# A FRAMEWORK FOR THE DEVELOPMENT OF MANUFACTURING SIMULATORS: TOWARDS NEW GENERATION OF SIMULATION SYSTEMS

V.V. Kryssanov*, V.A. Abramov*, H. Hibino[†], Y. Fukuda[‡]

*IACP, 5, Radio St., Vladivostok, 690041 Russia
[†]TRI of JSPMI, 1-1-12 Hachiman-cho, Higashikurume-shi, Tokyo, 203 Japan
[‡]Hosei University, 3-7-2 Kaginocho Koganei, Tokyo, 184 Japan

**ABSTRACT**

In this paper, an attempt is made to systematically discuss the development of simulation systems for manufacturing system design. General requirements on manufacturing simulators are formulated and a framework to address the requirements is suggested. Problems of information representation as an activity underlying simulation are considered. This is to form the necessary mathematical foundation for manufacturing simulations. The theoretical findings are explored through a pilot study. A conclusion about the suitability of the suggested approach to the development of simulation systems for manufacturing system design is made, and implications for future research are described.

## INTRODUCTION AND RESEARCH BACKGROUND

Recently, there is a widening belief that by applying simulation systems, it becomes possible to cope with the increasing technical, structural and organizational complexity of modern manufacturing enterprises while efficiently arranging production and administrative processes throughout the product life cycle. Manufacturing simulation systems help to realize and optimize the structure and properties of professional activities underlying technological processes, find a satisfying (if not the best) solution for a problem among the variety of feasible alternatives, and predict and analyze potential consequences (immediate as well as distant) of a candidate decision made concerning any of the product life cycle stages. Besides, simulation allows for the secure facilitation of operation and control of continuously complicated manufacturing systems.

One of the most promising manufacturing activities to employ simulation techniques and tools is manufacturing system design. There are three important phases when designing a complex production system [ISO, 1993]: 1) specification of system requirements, 2) system design, and 3) detailed (sub)system design. The first stage assumes definition of the system's desired capabilities and the expenses necessary to install the system, proceeding from the needed production volume, potentially available (material and manpower) resources, the company's intrinsic characteristics, and the market strategy chosen. The resulting specification is the basis for the description of the intended production process, material handling, and production management that is made during the second design phase. At this stage, all the major components of the planned system are evinced. The third phase stipulates detailing descriptions of the system components while optimizing the system efficiency. The appropriate machinery set, machine layout, and material handling route are then found, properties of the system components (subsystems) are coordinated, and the necessary administrative work is negotiated. At any of the named phases, simulation can be used to evaluate and possibly improve the human design decisions.

Despite the eventually onward march of the manufacturing system design process, the design phases are not distinctly sequential: there are many relationships and interconnections among the phases, and a design decision made at a later stage may cause the need to revise decisions made in the earlier stages. Moreover, the human design activity often has a cyclic iterated character when the design descriptions are complementarily formed in different ways at different scale abstractions over all three phases. Simulators used while designing complex manufacturing systems should then support the human processes of modeling the relevant pieces of the domain reality, coherently adjusting properties of the modeled objects and concepts in accordance with design decisions made at any stage of the formation of the system design. On the other hand, as any simulation models done are just once established approximations of the reality which cannot be fixed, these approximations need to be continuously validated and redesigned at timely intervals. Gaining knowledge about a manufacturing system is a never-ending process, and the simulation model development and the model validation are usually recurrently performed in a hierarchical manner for different levels of detail. Naturally, the easier it is to handle (edit, check, blend/split, etc.) simulation models adopted into a particular simulator, the higher the promise of the simulator's practical utility.

All these prerequisites of simulators, i.e., to properly map the domain under consideration and readily evolve the obtained domain model, dictate the following demands on next generation simulation systems for manufacturing system design:

• The need to support (and focus on) the process of conceptualizing (building a model of the reality) rather than to simply use (even if adjusting) completed models while the simulation is run. It is important to be able to construct simulation models not only consistently following the human process of conceptualizing but also explicitly specifying this process. The latter will considerably enhance the understandability and, ultimately, usability of simulation models, and reduce the efforts required to (re)use the models.
- The need of both validation and verification mechanisms as active, dedicated components of the simulators. Apart from general purpose verification (e.g., regarding design specifications, a standard representation, and/or reference models) and integrity checking functions which are required when a complex simulation model is built, it is necessary to be able to control (and/or realize) the verity of the model in respect to the reality. This is to provide for the effectiveness of the simulation and to promote the highest confidence for the simulation results.
- The necessity of a multi-level user (friendly) interface. The development of a simulator usually is an on-going process arbitrarily overlapping the process of the simulator use, and there are many potentially cooperative users of the simulator. Each of these users may adopt a distinct view of the intended simulation and therefore may need a particular interface to easily and purposefully interact with the system. For this, it is meaningful to give a user (developer, designer, or manager) the interface relevant to the user's professional needs and expertise.

Along with the highlighted points, as information technologies mature, there is the need to arrange mechanisms for sharing simulation information not only for the collaboration support within the frames of solving a particular task, but also for broad but systematic accumulation of the human experiences. This would reduce the effort for modeling, putting in (re)use simulation models previously built.

Neither of the above requirements is absolutely new nor unexplored, and there have been many works aimed at advancing simulation methods and tools to face the demands of actual manufacturing. Below, we will outline a few of the recent reports most relevant to our study.

Hierarchical reference models were suggested to guide the modeller while constructing simulation models (Mertins et al., 1997). Typical model structures, which are predefined generalized representations of the domain objects and concepts, are combined into a hierarchy layered by abstraction levels, and components of the resulting construction may be adapted to the needs of a particular simulation, reducing the abstraction. The authors suggested organizing a library for gathering simulation models built on the base of the predefined structures and to consecutively enlarge the hierarchy. These would ultimately facilitate the development of complex simulation models by allowing for reuse of existing structures, and stand for systematic verification of the reference models stored in the library.

In Osaki et al. (1997), a concept of open system architecture for a virtual shop floor was proposed to organize distributed simulation of manufacturing processes in an efficient and graphic manner. The modularization and distribution of information models were pointed to as key factors for the development of a virtual shop floor, and standardization of information representation and exchange was recognized as the activity crucial to practically accomplish the open system architecture for manufacturing simulators. The authors also realized the need of arranging an adaptable interface for different categories of the virtual shop floor users (e.g., simulation model developers and simulator end-users). A similar idea to individualize interfaces for simulator users in accordance with the users roles (for example, building models of manufacturing systems – for engineers, and adding new functionality to a simulator – for developers) was specifically discussed in an earlier work by Ball and Love (1995).

Arthur and Nance (1996) found that independent verification and validation of simulation models could enhance the models applicability when dealing with complex problems. Validation was characterized as an activity focused on building the right model or system, while verification was conceived as an effort to guarantee properly building the model or system. It was recommended in the study, to introduce an independent agent responsible for verification and validation as an active member into the modeling and simulation development process. Simulators should then be able to utilize the agent's expertise in both the simulation and the application domain.

Lingineni et al. (1996) presented a knowledge-based simulation model design tool. One of the basic ideas of that work was to facilitate the process of conceptualizing subject matter experts' knowledge about the reality by enabling the experts to describe the modeled entities and relations among them in a form that is natural to the domain. After the experts' descriptions are completed and verified, the intended simulation model can automatically be generated and thereupon run by an end-user, putting in service the corresponding functions of the developed tool.

In the literature, one may find a number of other reports treating problems similar to those pointed out above. However, it may be seen, that an overwhelming majority of these reports address a particular, situated aspect of the previously formulated demands on simulation systems, but no attempts to discuss their involvement have been made. Furthermore, many of the published studies on simulation techniques and tools although formulating the goal to introduce a new concept, methodology or even paradigm for manufacturing simulation, they obviously lack a fundamental vision of the problem and only abound in technical and realization details (e.g., see Kellert et al., 1997, just a recent example). The latter is revealed by the absence of a systematic theoretical background to simulation and the simulators developed which should be realized as an important challenge for research in manufacturing.

The focus of our study is on forming scientific principles for the development of a new generation of simulators for manufacturing system design. Some of these principles are approached in this paper. In the following section, we assign the scope, give the necessary assumptions, and provide a philosophy for our research. An information modeling framework for the next generation simulators is proposed in Section 3. In Section 4, we refer to a mathematical theory embraced in our study and discuss some formal aspects of the modeling and simulation development process. Then, Section 5 gives an account of a pilot study resulting in a prototype of an integrated environment for manufacturing system design simulation. Finally, in Section 6, we draw some conclusions from our experiences.

**BEHIND SIMULATION: INFORMATION REPRESENTATION**

We consider, that fundamental activities underlying computerized simulation are conceptualization and information representation, whereas any simulation model is an information model first. (Alternatively, we could at least assert that a simulation model is composed of information models.) By the ISO 10303 (STEP) standard, an information model is 'a formal model of a bounded set of facts, concepts or instructions to meet a specified requirement' (ISO, 1994). Such a model comes as a formalized outcome of conceptualization and is an abstract and simplified view of the domain that is to be represented for computer handling. We will regard conceptualization as a rational mental process of analyzing the reality by an individual agent or by a socially (environmentally, educationally, culturally, etc.) uniform group. Before thinking in detail about simulation and the simulators, we have to comprehend how to form an information (simulation) model. The latter evidently is one of the information representation questions.

In our study, we adapt the ideas described in Guarino (1994), distinguishing the five levels of information (knowledge) representation: logical, epistemological, ontological, conceptual, and linguistic. At the logical level, a representation sets to work with such primitives as relationships, predicates and functions, which usually have the standard formal semantics. This level is to assure logically supported formalization. The epistemological level is to offer information structuring primitives such as object, class, attribute, etc., as well as structural interrelationships among them (e.g., generalization, hierarchy, causality), not introducing any new semantics different from that which were fixed at the logical level. The ontological level serves to explicitly arrange a specification of ontological commitments, which assign the meaning for the structuring primitives of the epistemological level. This specification restricts the number of possible interpretations of the information (knowledge) expressed in a formalism by tying the formalism's constructions to the patterns and regularities of the objective reality. (We adopt the prevalent interpretation of the term 'ontology' as 'an explicit, partial account of a conceptualization,' while realizing conceptualization as it was settled above.) The conceptual level is to handle particular instances of domain concepts. Due to the cognitive nature of these instances, they may have only partially formalized semantics that are formed by inheriting semantics of the previous levels (if they were explicitly settled). Finally, the linguistic level delivers nouns, verbs and other parts that are directly used by people to express information.

A careful study of the available literature on manufacturing system design simulation has led us to the persuasion that considering information modeling, the focus of current research in this area is mostly on arranging information structuring constructions (generally or/and specifically) relevant to the simulation needs as well as on systematizing (sometimes, even standardizing) representations of the domain concepts (see Fig.1, where the circle A depicts the scope of the present research efforts).

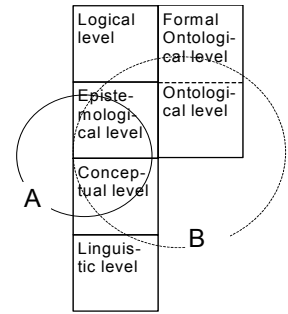

**Fig.1.** Levels of information representation: a simulation perspective

On the other hand, there is a widening awareness that when developing simulation models, it could be very profitable (e.g., for the reuse purposes) to put in service the ontological level of information representation (e.g., see Mertins et al., 1997, where the reference models may be understood as ontologies). Besides, the recently realized need to supplement simulation tools with mechanisms for knowledge acquisition and knowledge-based decision-making necessitates the intensive exploitation of the linguistic level to naturally deal with professional knowledge as much as possible (e.g., see Lingineni et al., 1996). Further, the increasing complexity of developed simulation models compels developers to intensively make use of the logical level of information representation to assure correct verification and integrity checking of the models. All these permit us to propose widening the scope of the information modeling and representation issues, which should be addressed in research on the development of manufacturing simulators, as it is shown in Fig.1 by the circle B.

In the following section, we introduce an architecture for an integrated software environment that is to satisfy the demands on simulators for manufacturing system design and contour a tool-kit that is necessary to thoroughly treat information (simulation) models. Below, are a few preliminary remarks for the tool-kit.

Considering conceptualization as a process, in our study the definition of ontology has a strong methodological flavor, and an ontology is dealt with as a description of the domain analysis made (or to be made) to form a given formal representation. The ontological level is also accountable for the social context of the representation use that, otherwise, would be lost. In our opinion, an ontology is not necessarily a fully consistent, provably correct and generic nor complete description. It can in some cases be a subjective and rather limited and situated vision of a phenomenon. In view of this, we will distinguish ontological descriptions, which are steady in respect to the reality, and place them at the formal ontological level (see Fig.1).

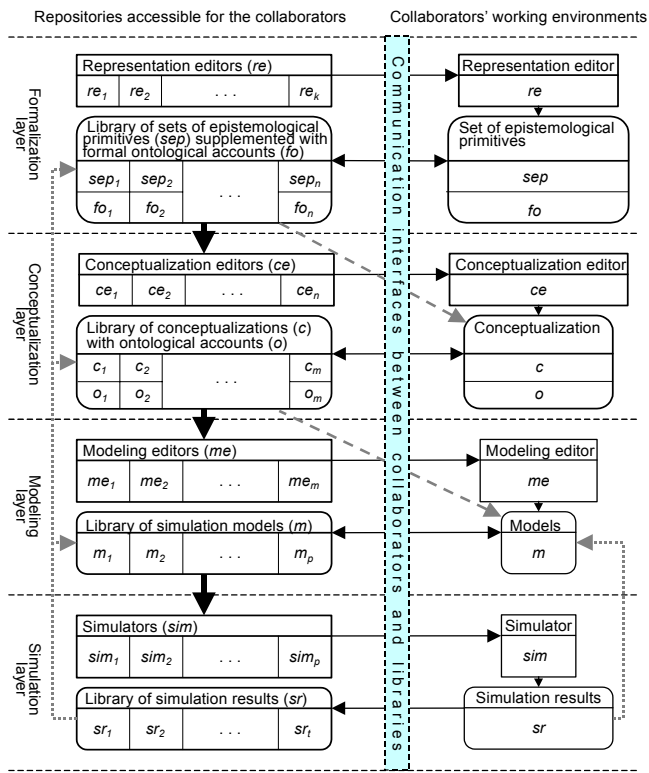

**Fig.2. Four layers of information modeling**

## A FRAMEWORK FOR THE DEVELOPMENT OF MANUFACTURING SIMULATORS

Finding an appropriate structuring of the information modeling activities is an important step to formulation of the scientific principles for the development of manufacturing simulators. Such a structuring should distinguish 'natural' (as if functionally decomposing the information modeling activities) layers of information modeling, not committing to any particular representation formalism, method, program, etc., and provide the necessary common basis for coordination of the modeling and simulation activities and tools. Having recognized these layers, the next step is to establish 'vertical' and 'horizontal' interfaces between them to make possible the necessary interactions. The vertical interfaces are to provide the continuity over different phases of the information modeling. The horizontal interfaces serve to promote collaboration within the same phase. The horizontal interrelations are also to allow for organizing common libraries of information representation means, conceptualizations, information models, etc., according to the layers structure. We have identified the following four layers of information modeling (Fig.2):

1) The formalization layer. The representation editors (which make use of the relevant knowledge from computer science and mathematics, such as logics, calculi, formal languages, and so on) are used to arrange collections of domain-specific epistemological-level primitives. These primitives are thoroughly imbued there with the specificity of the domain information (i.e., they get expressiveness necessary to deal with such domain- and simulation- specific aspects of information modeling as, for example, topological, causal, temporal relationships among the domain entities). A formal ontological account on the primitives meaning is made by specifying a (partially) formalized set of rules for mapping the domain entities to the suitable epistemological primitives. The creation of this account is of great importance, as the ontological specifications explicitly made, always facilitate the processes of conceptualization, verification and validation (see Guarino, 1994). It should be noted however, that the formalization layer is not to unify the set of epistemological primitives and/or rules but rather to build many specific sets of such primitives and create the possibility to compare and relate these sets. The latter is achieved by installing the library of epistemological primitives sets and the development of the horizontal interfaces between the users (who make up and utilize the primitives).

2) The conceptualization layer. Domain-specific conceptual structures learned and/or composed by people are mapped to a set of epistemological primitives under restrictions of the relevant rules which were settled earlier for mapping the domain entities to the epistemological-level primitives. To support this process, the conceptualization editors are built so that each editor is to handle representations written in terms of one set of epistemological primitives. These tools are also to prepare ontological accounts of the conceptualizations. Generally, the conceptualization layer is to specify particular conceptualizations, compare and relate them. The library of conceptualizations and the horizontal interfaces are also needed at this layer to support the collaboration.

3) The modeling layer. Conceptual structures from the previous layer are used to make up information models. The models are constructed, manipulated, related, filed to the simulation models library with the modeling editors and transformed to executable models (adopted in simulators) through the vertical interface. Verification of the information models is possible by making use of the restrictions of the ontological accounts. This prevents improper use of the epistemological-level primitives and conceptualizations completed, but only partially assures validity of the information models built due to the incomplete character of the ontological commitments.

4) The simulation layer. This is to immediately utilize information models obtained, applying them for simulation.

There may be several concurrent users of the resulted simulation models who access a common collection of simulators. Simulation results can be gathered in the library and further used as a feedback to the modeling activity as well as to facilitate validation of the simulation models over all the layers of information modeling.

It can be seen, that the suggested framework has a plain arrangement to provide networked, distributed and concurrent collaboration aimed at information modeling and simulation. However, to practically establish the introduced layers structure and the embedded mechanisms, it is necessary to have a strong mathematical foundation capable of managing the whole variety of information representation constructions in a controllable, technologically maintainable and unified manner. In the next section, we point to a mathematical apparatus adequate for the information modeling and simulation needs.

## ON FORMALIZATION

We will consider, that an information (simulation) model is composed of a collection of information entities. An entity can represent an object (material or conceptual), an operation (action, function, or role) that is defined on objects, a situation (event, task, or scenario) that reflects a context of objects use, or a process that shows an arrangement of events. An information entity can have attributes describing the entity properties. An attribute can be another entity or a data parameter. The latter consists of data descriptions such as numbers, textual characters, mathematical signs, etc. The relationships among attributes establish the structure of the information entity, and the relationships among entities constitute the structure of the information model. An information (simulation) model may also be supplemented by descriptions of application rules - constraints, specialization and generalization rules, heuristics, and the like, which give an appropriateness, behavioral and utilization context for the given model.

To make the above image of simulation models formal, in our research we adopt the notation of sorted (indexed) sets. More precisely, we use the theory of algebraic specification that does provide powerful resources to handle simulation models over all the levels of information representation (for basics of algebraic specification, see Goguen and Malcolm, 1996). Assuming that the intended realization of the representation (e.g., with a programming language) permits dealing with information (simulation) models as strings of symbols (that, of course, does not interdict graphic and 'visual' languages, animation, etc.), we introduce the following principles of formalization.

Let $\Sigma$ be a set of symbols, which are unique names assigned to the domain objects, object data parameters, and relationships among the objects and parameters. All the symbols are arranged into sorts so that every symbol can have many different sorts. The sorts are grouped with sort sets, not necessarily disjoint, while the sort sets may belong to a sort superset. Every sort set is ordered by a 'sort-subsort' hierarchy. In turn, the sort sets may be ranked. All the orderings are partial and mutually independent.

We will consider such a set $\Sigma$ as a terminological alphabet of the domain under consideration. Then, an information entity $S$ of the domain can be specified as $S: {}^mF(A,B)$ if $C$. Here, $A \in \Sigma^*$ is a set of attributes of $S$ (constants are allowed), and $B \in \Sigma^+$ is the sort of $S$ ($\Sigma^*$ denotes the set of all sequences of strings that are composed of zero or more symbols of $\Sigma$, and $\Sigma^+$ denotes the set of all sequences of strings that are composed of one or more symbols of $\Sigma$); $C$ is a set of predicates defined on $\Sigma$ as application rules for $S$; m is a mode of a functor $F$. This functor establishes a relation on $A$ and $B$ and includes at least one function $f: A' \to B$ such that whenever $f \subseteq F$, $A' \in \Sigma^*$ and $A \cap A' \neq \emptyset$; m is to specify a way in which $A$ and $A'$ are related.

It can be seen, that the use of the described principles for formalizing information permits us to emphasize the structure of the domain information representation rather than 'ad hoc' chosen properties of an individual entity during conceptualization. The sort sets can serve to functionally layer the intended (formal) representation by the underlying human activities, such as notation and software development, information acquisition, information representation and processing, manufacturing system design, modeling and simulation. The factors affecting a conceptualization of the domain and the resulting representation of a simulation model (such as domain analysis methodology, abstraction, domain-specificity, etc.) can also be specified and, then, interpreted, managing sorts assigned to the model and/or defining properties of the function $f$. As an illustration of the latter proposition, let us consider possible ways to treat abstraction and view (aspect, perspective) which play important roles in the modeling and simulation development process.

We perceive abstraction as the mapping from one representation (of a problem) to another which preserves certain desirable properties but reduces complexity (see Giunchiglia and Walsh, 1992). It is a well-known fact, that abstraction has great implications for designing manufacturing systems and for developing and using manufacturing system models as well. In particular, abstraction by granularity of description is often tackled in research on manufacturing system simulation (e.g., Hibino et al., 1997; Mertins et al., 1997). However, it should be noted that while there are many reports on the application of completed information constructions differentiated by different kind abstractions for the simulation purposes, the notion of abstraction itself is still very intuitive and not formalized. This gives rise to many complications when one tries to understand and/or (re)use the models obtained by different authors as the authors' conceptualizations have not been explicitly specified. In the framework of our approach, abstraction by granularity of description can be manipulated as follows.

Given a representation of a concept $S: {}^mF(A,B)$ if $C$, $\Sigma$ is a terminological alphabet of the domain under consideration, and $T$ a sort set for $\Sigma$. Then, a more abstract representation of the same concept can be specified as $S: {}^mF(A^1,B)$ if $C$, where $A^1$ is a set of attributes such that for every $a \in A_t$, $t \in T$, there is $b \in A^1_{t1}$, $t1 \in T$ and $t \leq t1$. Similarly, a refined representation of the concept can be specified as $S: {}^mF(A^2,B)$ if $C$, where $A^2$ is a set of attributes such that for every $c \in A^2_{t2}$, $t2 \in T$, there is $a \in A_t$, $t \in T$, and $t2 \leq t$. It may be seen, that explicitly defining m a mode of the functor $F$ as a function of the abstraction, it becomes possible to form an ontological account of the conceptualization while creating representations of a domain object through various levels of the abstraction.

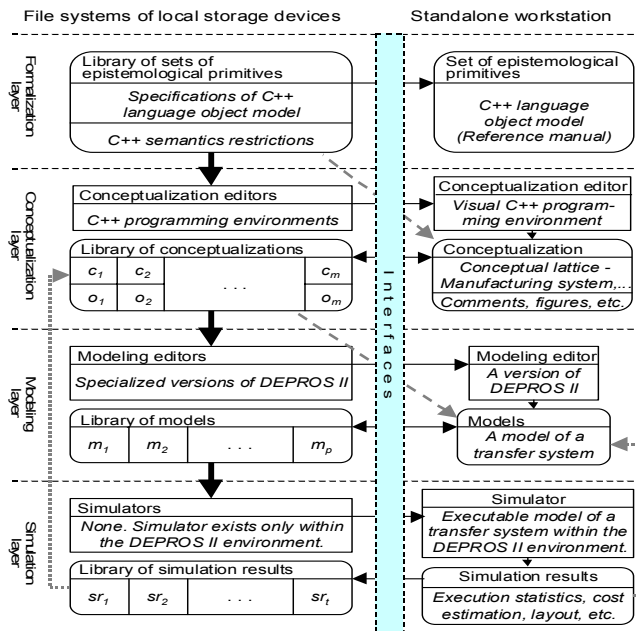

**Fig.3.** The pilot study simulation environment
(see Fig.2 for designations)

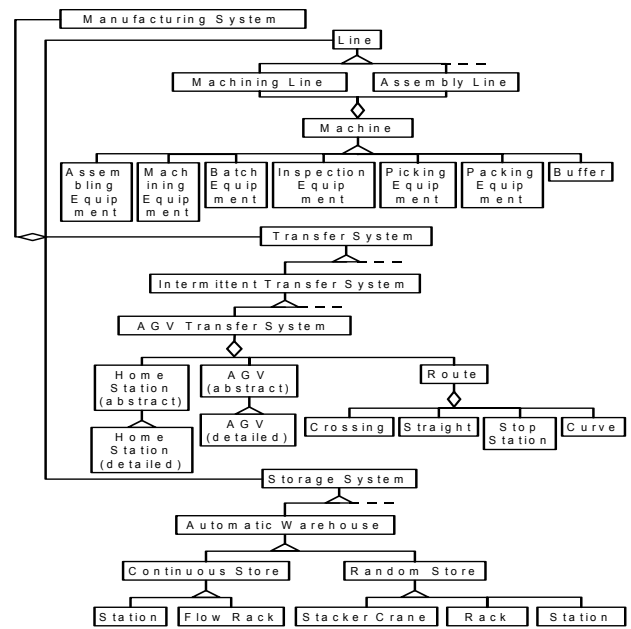

**Fig.4.** A fragment of a conceptual lattice for
modeling manufacturing systems

A different critical factor influencing conceptualization of the domain under consideration is a view of the domain that reveals as context-dependency, the resulted representation. For example, considering the simulation of a manufacturing system, one may be interested in estimating the operating costs of the system, while another - in adjusting the system logistics (however, a distinct view does not necessarily imply a distinct user). Having the same modeled phenomenon, there would be many diverse views resulting in many different conceptualizations and, ultimately, many separated models of the phenomenon. An attempt to merge and integrate the models will inevitably lead to the substantially unmanageable multipurpose simulation model with numerous actions, goals, relations and contradictions among them. This problem can hardly be solved by technical means, but could be controlled at the level of formalization.

One of the feasible ways to cope with the complexity of a multipurpose simulation model is to handle a different view of the domain with a distinct sort set. Then, the sort sets system is to put the model into context, and properties of a specified domain object are essentially determined by the sort assigned to the object. As the resulting representations could be accumulated in the course of time, the sort sets system could be simplified by collating the existing stable relationships among the sort sets and ranking the sorts.

Other potentialities of the suggested principles of formalization can be found in Kryssanov et al. (1997).

## A PILOT STUDY

In order to explore the suitability of the framework for the development of manufacturing simulators as an aid to advance simulator architecture, and to verify the functionality and utility to satisfy the demands on simulation systems for manufacturing system design, a pilot study has been performed. Taking DEPROS II, an earlier developed simulator for production system design (see Hibino et al., 1997) as a prototype, we have changed its architecture, adding new modules to enhance the simulator functionality and, in this way, assembled an integrated environment for manufacturing system design simulation. Using the newly constructed environment, we have developed a simulation model for a manufacturing system, verified it, and made a simulation run. The simulation results obtained have been used to validate (and are used to improve) the simulation model as well as the underlying information representation constructions of the epistemological, ontological and conceptual levels.

Fig.3 depicts the structure of the simulation environment used in the pilot study (also, see the designations in Fig.2). For the environment implementation, Visual C++ and Intelligent Pad (Intelligent Pad, 1996) have been used.

The modeled manufacturing system consists of two machining lines, one assembly line, one automatic warehouse system and an intermittent transfer system using Automated Guided Vehicles (AGVs). A conceptual lattice for the model, which is the result of conceptualization made in the study, is partially shown in Fig.4. Together with the ontological commitments, this structure aids understanding the meaning of the representations made and so facilitates the model reuse.

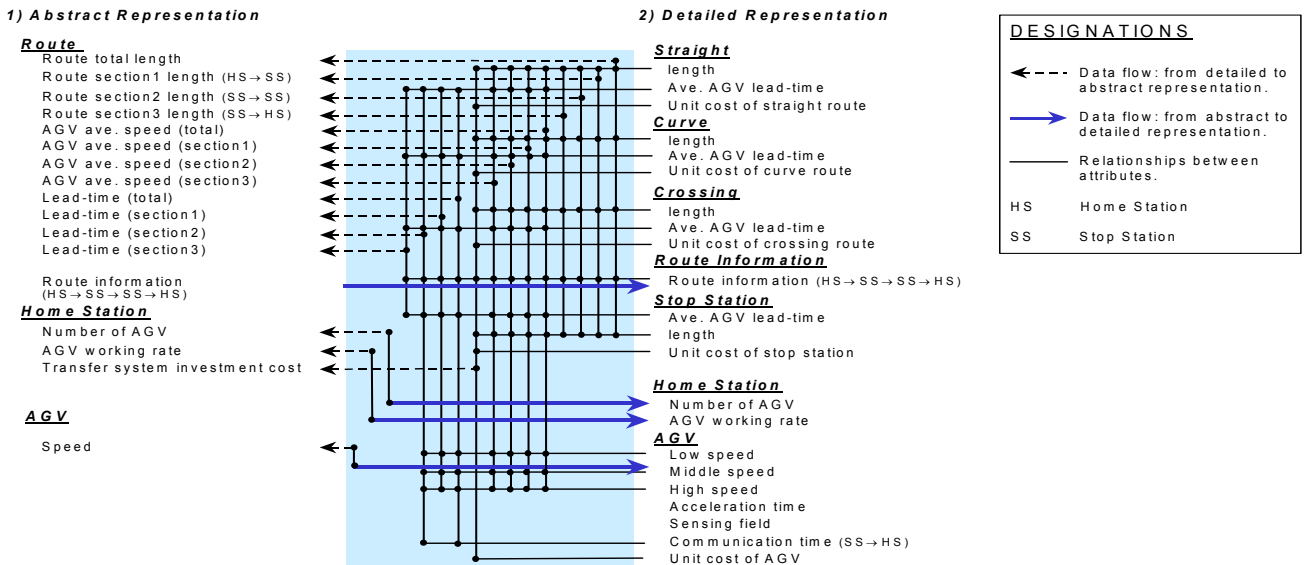

**Fig.5.** Representation of an information model on two different levels of the abstraction

Fig.5 presents an example of handling the abstraction with the developed simulator (see also Fig.6). There are two modes for the transfer system model: abstract (1) and detailed (2). In the abstract mode, the model includes a route model, a home station model, and an AGV model. The detailed representation of (for instance) the route model is composed of models of straight tracks, curves, crossings, stop stations, and route data. Each of the models has attributes, and the relationships between the attributes in the abstract and detailed modes are shown in the figure. In the same manner, Fig.5 depicts the mappings between abstract and detailed representations of the home station and AGV.

The process of developing the simulation model involved the following actions. First, (the models of) the machining line, the assembly line, the automatic warehouse system, and the transfer system were arranged (in the abstract mode) according to the intended system specifications, and the capacity necessary for the transfer system was estimated. Then, the performance and capabilities of the manufacturing system were determined. At the next stage, (the models of) AGVs, the routes and the home stations were ordered (in the detailed mode), and the evaluations for the transfer system were made. Finally, all the simulation model components were verified, while the abstract and detailed representations of the components were coordinated. The completed simulation model was put in the model library of the simulator. An example of a simulation run is shown in Fig.6.

From experiences drawn from the pilot study, it was considered that the suggested framework was easily applied to the existing simulation system. The resulting re-designed simulator enables quick and simple definition and the running of manufacturing system simulation models over different levels of abstraction. The separated interfaces for the developers and designers facilitate the users' interaction with the system, while the access to the information at the conceptual level of the model representation and the ability of (not even formally) dealing with the underlying ontological information assist proper understanding of the formalized constructions and hence promote accurate reuse of the simulation models accumulated in the model library. Another finding of the pilot study was that the suggested architecture for the simulation environment allows for efficiently organizing verification and validation of simulation models and the models' constituents as well. Although the implemented systems have no dedicated facilities to support validation, it could to a certain extent be done with the modeling editor, collating the simulation results and the corresponding phenomenon from reality and, then, making the necessary adjustments of the model representation.

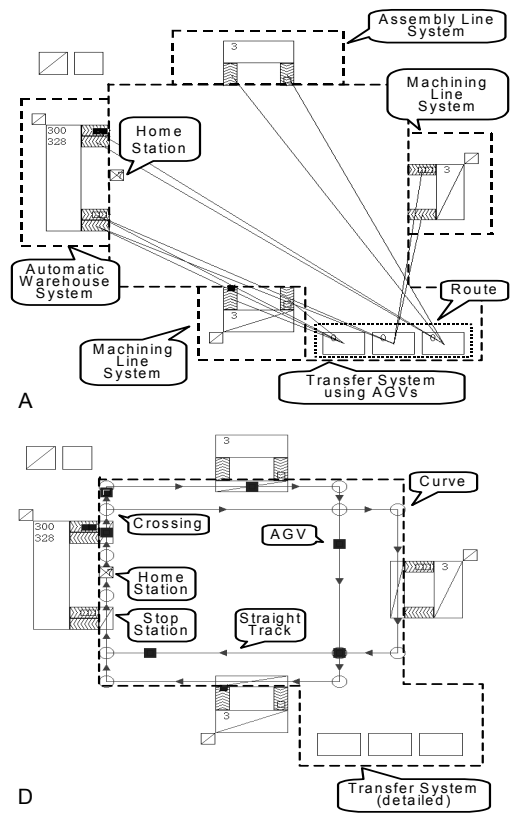

**Fig.6.** A sample run of the simulator
(A – in the abstract mode; D – in the detailed mode)

In general, the results obtained through the pilot study strengthened our supposition that the introduced framework for the development of manufacturing simulators satisfies the demands on simulation systems for manufacturing.

**CONCLUDING REMARKS**

Developing computer simulators is continuously increasing for manufacturing. Today's competitive market requires building manufacturing systems on larger and more complicated scales than ever before, and simulation is often used to organize effective support while designing such systems. However, practice shows us that the simulators presently in use are rather difficult to operate, inflexible, and insufficient to support the human activities throughout the manufacturing system design cycle. Simulation models developed are usually difficult to classify, share and reuse due to the lack of a common representation layer. The latter leads to mutual incompatibility of the simulators, both technically and conceptually. Overall, current efforts on simulation in manufacturing are not systematized, and there is an obvious deficiency of fundamental research on the subject.

In this study, we have made an attempt to scientifically approach the problem of manufacturing simulators development. Through the review of the available literature, the demands on simulation systems for manufacturing system design were formulated, and the general information modeling framework for a new generation of manufacturing simulators was introduced. Information representation was pointed to as a crucial activity while building simulation models. In this paper, we have also considered the important philosophical and mathematical issues of information representation. These theoretical issues help to clearly show the strengths and the weaknesses of the existing simulation applications and to detect the directions for enhancing the functionality, flexibility, and usability of manufacturing simulators. To check the appropriateness of our ideas, the pilot study was conducted. Taking an earlier developed simulator, its architecture was changed so that the simulator received the new functionality and became easier to operate and to access. The results of the pilot study permit us to consider the suggested approach as promising for further elaboration.

In future work, we intend to continue research on scientific principles for a new generation of simulators for manufacturing system design. The next points for investigation will be validation of a simulation model over all the levels of information representation and networking for the simulation environment. Besides, we plan to improve the simulation environment to manage multipurpose simulation models.

**Acknowledgment**

V. Kryssanov and V. Abramov gratefully acknowledge the financial support of the New Energy and Industrial Technology Development Organization (NEDO), Japan, under the research project 'Development of the Production Design Technology for Machining.' H. Hibino and Y. Fukuda acknowledge the receipt of the IMS research programme award 'IMS9710: Modeling and Simulation Environments for Design, Planning and Operation of Globally Distributed Enterprises (MISSION).'